\documentstyle[aps,amsfonts,epsfig]{revtex}   

\newcommand{\beq}{\begin{eqnarray}}
\newcommand{\eeq}{\end{eqnarray}}

\newcommand{\pardis}{\langle \mu \rangle}

\begin{document}     

\twocolumn[\hsize\textwidth\columnwidth\hsize\csname
@twocolumnfalse\endcsname
  
\draft

\title{Color confinement and dual superconductivity in full QCD}
 
\author{J.M. Carmona$^{a,1}$, M. D'Elia$^{b,2}$, L. Del Debbio$^{c,d,3}$, 
A. Di Giacomo$^{c,d,4}$, B. Lucini$^{e,5}$, G. Paffuti$^{c,d,6}$}

\address{$^a$ Departamento de F\'{\i}sica Te\'orica, Universidad de Zaragoza, 
50007 Zaragoza, Spain}
\address{$^b$  Dipartimento di Fisica dell'Universit\`a di Genova and INFN,
Sezione di Genova, Via Dodecaneso 33, I-16146 Genova, Italy}
 \address{$^c$ Dipartimento di Fisica dell'Universit\`a, Via Buonarroti
2 Ed. C, I-56127 Pisa, Italy}
\address{$^d$ INFN sezione di Pisa, Via Vecchia Livornese 1291, I-56010
S. Piero a Grado (Pi), Italy}    
\address{$^e$ Theoretical Physics, University of Oxford,
1 Keble Road, OX1 3NP Oxford, UK}
\address{$^1$ e-mail address: jcarmona@posta.unizar.es}
\address{$^2$ e-mail address: delia@ge.infn.it}
\address{$^3$ e-mail address: ldd@df.unipi.it}
\address{$^4$ e-mail address: digiaco@df.unipi.it}
\address{$^5$ e-mail address: lucini@thphys.ox.ac.uk}
\address{$^6$ e-mail address: paffuti@df.unipi.it}
        
\maketitle

\begin{abstract}
We report on evidence that confinement is related to dual superconductivity 
of the vacuum in full QCD, as in quenched QCD. The vacuum is a dual superconductor
in the confining phase, whilst the $U(1)$ magnetic symmetry is realized 
{\em \`a la} Wigner in the deconfined phase. \end{abstract}
\pacs{PACS numbers: 11.15.Ha, 12.38.Aw, 14.80.Hv, 64.60.Cn}
]

\section{Introduction}              

In a series of papers \cite{artsu2,artsu3,artran}, 
which we shall refer to as I,II,III 
respectively, we have demonstrated by numerical simulations that quenched QCD 
vacuum is a dual superconductor in the confining phase, and goes to normal 
state at the deconfinement transition (see also Ref.~\cite{bari}). 
More precisely we have constructed
an operator $\mu$ carrying magnetic charge and we have measured its vacuum expectation
value $\langle \mu \rangle$. For $T < T_c$ 
$ \langle \mu \rangle \neq 0$, for $T > T_c$ $ \langle \mu \rangle = 0$
and, approaching $T_c$, $\langle \mu \rangle \simeq (1 - T/T_c)^\delta$
 ($\delta = 0.50(3)$).
Magnetic charge is defined by a procedure called Abelian projection:
it associates $(N_c - 1)$ $U(1)$ magnetic symmetries to any operator $\phi$ in 
the adjoint representation~\cite{thooft81}. A priori magnetic symmetries corresponding
to different Abelian projections (different choices of $\phi$) are independent.
In I, II, III we have shown that the behaviour of $\langle \mu \rangle$,
including the value of the index $\delta$, is independent of the Abelian
projection.

There is general agreement on the order-disorder nature of the deconfining
transition in the quenched case. The popular order parameter is the Polyakov 
line $\langle L \rangle$; the symmetry
involved is $Z_N$. Alternatively the dual ('t Hooft) line 
$\langle \tilde{L} \rangle$~\cite{thooft78} can be used as a disorder parameter
(order parameter of the disordered phase) corresponding to the dual
 $\tilde{Z}_N$ symmetry.
Our $\langle \mu \rangle$ is also a good disorder parameter, and in fact 
it coincides numerically with $\langle \tilde{L} \rangle$~\cite{vort1,vort2}.

In full QCD, {\em i.e.} in the presence of dynamical quarks, the situation
is less clear. $Z_N$ and $\tilde{Z}_N$ symmetries are explicitely broken
by the very presence of the quarks. At zero quark mass there is a 
phase transition at some $T_c$ involving chiral symmetry:
for $T < T_c$ chiral symmetry is spontaneously broken, the pseudoscalar
octet being the Goldstone particles, and for $T > T_c$ it is restored.
Quark masses do break chiral symmetry explicitely. It is not clear 
theoretically what the chiral transition has to do with the
deconfinement transition. However, the susceptibilities
of different quantities (the Polyakov line $\langle L \rangle$, 
the chiral condensate) have been measured, 
and all of them have a maximum at the same value
$T_c ( m_q)$ for any value of $m_q$. Above a certain value of
$m_q$ ($m_q > 3$ GeV) the transition is first order, as in the quenched case, 
and $\langle L \rangle$ still works as an order parameter. At 
$m_q \sim 0$ the transition is presumably second order. At intermediate
values the susceptibilities which have been considered
show a maximum at $T_c$, but it does not go large at increasing 
volume. The indication is then that there is no transition but only a 
crossover.

A natural question is then if dual superconductivity is a symmetry for the 
transition in full QCD as it is in the quenched case.
In the spirit of the $N_c \to \infty$ limit, one would expect 
that the mechanism of confinement be the same as in quenched QCD, 
the idea being that the structure of the theory is the same as that in 
the limit $N_c \to \infty$ at $g^2 N_c = \lambda$ fixed: at finite
$N_c$ small differences are expected with respect to the limiting case. 
Quark loops are nonleading in the expansion. The mechanism of confinement
should be approximately $N_c$ independent and the same with and without dynamical 
quarks.

The disorder parameter $\langle \mu \rangle$ can be constructed in full
QCD exactly in the same way as in the quenched case (see Section 2).
At a given temperature $T$, $\langle \mu \rangle$ has to be computed
in the infinite volume limit. We have investigated the region
$T < T_c$, where we find 
\beq
\lim_{V\to\infty}\langle \mu \rangle \neq 0 
\eeq
and $T > T_c$, where we find
\beq
\lim_{V\to\infty}\langle \mu \rangle = 0
\label{ht}
\eeq
as will be shown in detail below. Notice that the limit Eq. (\ref{ht})
is not within errors but exact. Indeed we measure, instead of 
$\langle \mu \rangle$, the quantity 
$\rho = \frac{d}{d \beta} \ln \langle \mu \rangle$, and we find that
it tends to $- \infty$ as $\rho = - k N_s + k'$ ($k > 0$) as the
spatial size of the sample $N_s \to \infty$.

The finite size scaling analysis in the critical region
is under investigation, to study the nature and the order of the transition.

\section{Disorder parameter}

The operator $\mu$ is defined in full QCD exactly in the same way as in the
quenched theory \cite{artsu2,artsu3,artran}
\beq
\label{defmu}
\langle \mu \rangle = \frac{\tilde{Z}}{Z} \; ,\nonumber \\
Z = \int \left( {\cal D}U \right)  e^{-\beta S} \; ,\nonumber \\   
\tilde{Z} = \int \left( {\cal D}U \right)  e^{-\beta \tilde{S}} \; .
\eeq
$\tilde{Z}$ is obtained from $Z$ by changing the action in 
the time slice $x_0$, $S \to \tilde{S} = S + \Delta S$.
In the Abelian projected gauge the plaquettes
\beq
&&\Pi_{i0} (\vec{x},x_0) = \nonumber \\
&& = U_i (\vec{x},x_0) U_0 (x + \hat{\imath},x_0)
U_i^\dagger(\vec{x},x_0 + \hat{0}) U_0^\dagger (\vec{x},x_0)
\eeq
are changed by substituting
\beq
U_i(\vec{x},x_0) \to \tilde{U}_i(\vec{x},x_0) \equiv
U_i(\vec{x},x_0) e^{i T b_i (\vec{x} - \vec{y})}
\eeq
where $\vec{b}  (\vec{x} - \vec{y})$ is the vector potential of a 
monopole configuration centered at $\vec{y}$ in the gauge
$\vec{\nabla} \vec{b} = 0$, and $T$ is the diagonal 
gauge group generator corresponding to the monopole species chosen.
In SU(2) $T = \sigma_3 /2$, in SU(3) $T = \lambda_3/2$ or
$(\sqrt{3} \lambda_8 - \lambda_3)/2$. In the generic SU($N$) case
the procedure is explained in Ref.~\cite{sun}. 
Unlike the $Z_N$ centre symmetry, 
the $U(1)$ magnetic symmetry defined after Abelian projection is 
a good symmetry also in presence of dynamical fermions.
It can be shown that, as in the quenched case, 
$\mu$ adds to any configuration the monopole configuration
$\vec{b} (\vec{x} - \vec{y})$. If the magnetic symmetry is realized
{\em \`a la} Wigner, $\langle \mu \rangle = 0$ if $\mu$ carries
non zero net magnetic charge. Then $\langle \mu \rangle \neq 0$
means Higgs breaking of the U(1) symmetry. 
Therefore $\pardis$ can be a correct disorder parameter
for the transition to dual superconductivity also in full QCD.

\section{Numerical results}

We have measured $\pardis$ with two flavours of 
degenerate staggered fermions on $N_s^3 \times 4$ lattices, with
different values of $N_s$ ($N_s = $ 12,16,32) 
and of the bare quark mass $m_q$. 
In particular we have chosen, in the transition region, 
to vary the temperature, $T = 1/( N_t a(\beta, m_q))$, 
moving in the $(\beta, m_q)$ plane while keeping a fixed value of 
$m_\pi/m_\rho$. To do this
and to extract the physical scale we have used fits to the 
$m_\rho$ and $m_\pi$ masses published in~\cite{blum}. 
We present here results obtained at $m_\pi/m_\rho \simeq 0.505$:
in this case, at $N_t = 4$, the $\beta$ corresponding to the 
transition is approximately $\beta_c \sim 5.35$~\cite{jlqcd}.
Preliminary results have been already presented in~\cite{latt01}.

Instead of $\pardis$ we measure the quantity
\beq
\rho = \frac{d}{d \beta} \ln \langle \mu \rangle \; .
\eeq
It follows from Eq. (\ref{defmu}) that
\beq
\rho = \langle S \rangle_S -  \langle \tilde{S} \rangle_{\tilde{S}} \; ,
\label{rhoferm}
\eeq
the subscript meaning the action by which the average is performed. 
In terms of $\rho$
\beq
\label{mufromrho}
\pardis = \exp\left(\int_0^{\beta} \rho(\beta^{\prime})\mbox{d}\beta^{\prime}\right) \; .
\eeq
A drop of $\pardis$ at the phase transition corresponds to a strong
negative peak of $\rho$.

We have used the R version of  the HMC algorithm for our 
simulations~\cite{gott87}.
Some technical complications arise in the computation
of the second term on the right hand side of Eq. (\ref{rhoferm}).
In the evaluation of 
$\langle \tilde{S} \rangle_{\tilde{S}}$, $C^{\star}$-periodic boundary
conditions in time direction have to be used for the gauge fields and this 
requires $C^{\star}$ boundary conditions in
temporal direction also for fermionic variables 
(in addition to the usual antiperiodic ones), in order to ensure gauge 
invariance of the fermionic determinant.
This implies relevant changes in the formulation
and implementation of the HMC algorithm which are explained in detail
in Ref.~\cite{cstar}.

We have chosen the Polyakov line as the local adjoint operator which 
defines the Abelian projection. Actually, calling $L(\vec{x},x_0)$ 
the Polyakov  line starting at point $(\vec{x},x_0)$, 
the Abelian projection is defined by the operator 
$L(\vec{x},x_0) L^\star(\vec{x},x_0)$, which transforms 
in the adjoint representation when using $C^\star$ boundary 
conditions.

The use of a modified gauge action also implies changes in the 
molecular dynamics equations. One has to maintain 
the modified hamiltonian containing $\tilde{S}$ constant. 
A change in any temporal link indeed induces
a change in $L(\vec{x},x_0)$ and hence in the Abelian projection 
defining the monopole field. Therefore the dependence of $\tilde{S}$ on 
temporal links is non trivial and the equations of motion
for the temporal momenta become more complicated. 

Fig.~1 shows $\rho$ for a $32^3 \times 4$ lattice, and the chiral condensate 
as a function of $\beta$. The negative peak of $\rho$ is clearly
at the same value of $\beta$ where $\langle \bar{\psi} \psi \rangle$
drops to zero.

Fig.~2 shows the plot of $\rho$ for different spatial sizes
$N_s$. For larger lattices the peak becomes
deeper and the value of $\rho$ at high $\beta$ lower. 

\begin{figure}[]
\vspace{0.5cm}
\centerline{\epsfig{figure=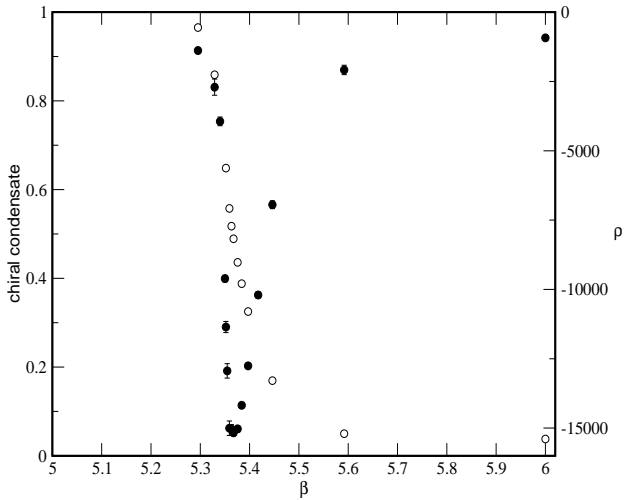,angle=0,width=82mm,height=66mm}}
\vskip 0.86cm
\caption{Chiral condensate (open circles) and $\rho$ (filled circles)
on the $32^3 \times 4$ lattice.}
\label{fig:rhovspsi}
\end{figure}

An analysis of $\rho$ at large $\beta$'s as a function of
$N_s$ is shown in Fig.~3, for different masses of the staggered
fermions used in the simulations.
For net magnetic charge $\neq 0$
\beq
\label{rhowk}
\rho \simeq -k N_s + k' \;\;\;\;\; (k > 0)
\eeq
and is practically independent of the quark mass within errors.
For net charge zero (e.g. monopole-antimonopole pair)
$\rho$ stays constant at large $N_s$. Going back to Eq. (\ref{mufromrho})
this means that $\pardis$ is strictly zero in the infinite
volume limit for non zero  magnetic charge, and can be
$\neq 0$ for excitations with zero net magnetic charge.
This statement is based on the analysis of many different excitations
with different magnetic charges, and Fig.~3  is only an example.
The magnetic symmetry is therefore realized {\em \`a la} Wigner
for $T > T_c$ and the Hilbert space is superselected.
Notice that:

(1) $\pardis$ can only be strictly zero in the infinite volume limit
 (Lee-Yang theorem~\cite{lee}).

(2) If we were measuring $\pardis$ directly we would find zero within
     large errors. Looking instead at $\rho$ we can unambigously
    check Eq. (\ref{rhowk}), which means that $\pardis$ is strictly zero
    as $N_s \to \infty$.

For $T < T_c$ $\pardis \neq 0$ if $\rho$ stays constant
and finite with increasing volume. This is what indeed happens
as shown in Fig.~4. Nothing spectacular can happen at larger
volumes, since no larger length scale  exists in the system.

Around $T_c$ a finite size scaling analysis is required to get information
on the order of the transition as well as to measure the critical
indices. The problem is more  complicated than in the quenched case,
since an extra scale, the quark mass, is present. The program is on the
way on a set of APEmille machines.
Some qualitative features are shown in Fig.~2.

\begin{figure}[h]
\centerline{\epsfig{figure=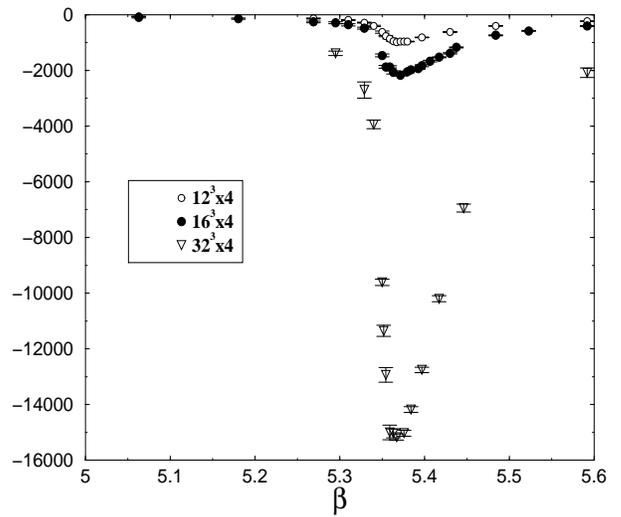,angle=0,width=80mm,height=70mm}}
\vskip 0.5cm
\caption{Behaviour of $\rho$ around the phase transition at various 
lattice sizes.}
\label{fig:transition}
\end{figure}

\begin{figure}[h]
\centerline{\epsfig{figure=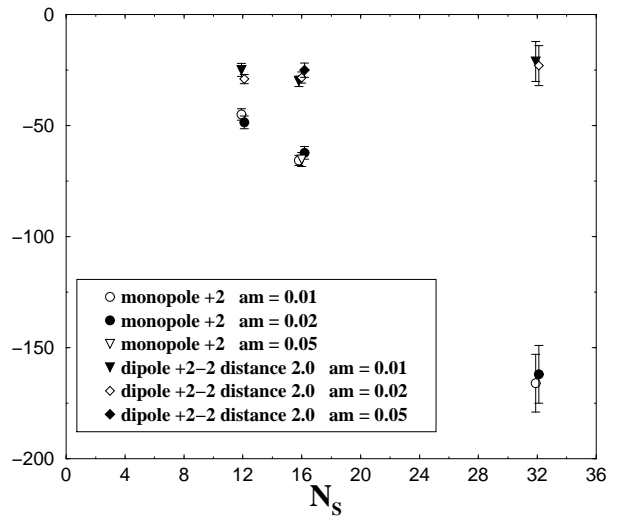,angle=0,width=80mm,height=70mm}}
\vskip 0.5cm
\caption{Weak coupling behaviour of $\rho$ at various lattice sizes.}
\label{fig:weakc}
\end{figure}     

\section{Conclusions}

The preliminary data reported in this paper contain enough
information to state that dual superconductivity is at work
as a confinement mechanism in QCD with dynamical quarks,
in the same way as in the quenched theory~[I,II,III].
For $T > T_c$ the Hilbert space is superselected with respect to
magnetic charge, for $T < T_c$ the symmetry is Higgs broken.

Dependence of the disorder parameter on the choice of the 
Abelian projection and nature of the transition are under investigation.

\begin{figure}[h]
\centerline{\epsfig{figure=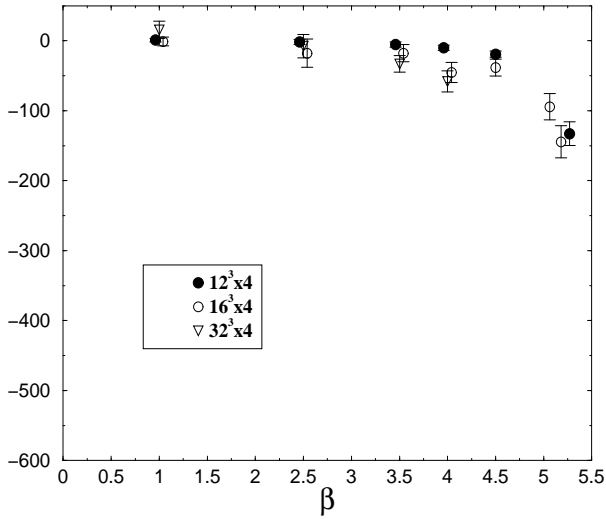,angle=0,width=80mm,height=70mm}}
\vskip 0.5cm
\caption{Strong coupling behaviour of $\rho$ at various lattice sizes
and $am = 0.1335$.}
\label{fig:strongc}
\end{figure}

\section*{Acknowledgements}
This work has been partially supported by MIUR and by EU contract 
No. FMRX-CT97-0122. BL is supported by the Marie Curie Fellowship 
No. HPMF-CT-2001-01131.

\end{document}